\documentclass[aoas]{imsart}

\RequirePackage{amsthm,amsmath,amsfonts,amssymb}
\usepackage{threeparttable}
\usepackage{tabularray}
\usepackage{booktabs}

\usepackage{fix-cm}
\usepackage{lmodern}

\usepackage{graphicx}
\usepackage{xcolor} 
\usepackage{hyperref}
\hypersetup{
    colorlinks=true,
    linkcolor=blue,
    citecolor=blue,
    urlcolor=blue
}

\RequirePackage[authoryear]{natbib}

\newcommand{\iid}{\stackrel{\rm iid}{\sim}}

\startlocaldefs
\theoremstyle{plain}

\theoremstyle{definition}

\endlocaldefs

\begin{document}

\begin{frontmatter}
\title{Linear Models, Variable Selection, Artificial Intelligence}
\runtitle{Variable Selection via AI}

\begin{aug}
\author[A]{\fnms{Riyadh}~\snm{Alrawkan} \ead[label=e1]{alrawkanr@vcu.edu}},
\author[A]{\fnms{Edward}~\snm{Boone}\ead[label=e2]{elboone@vcu.edu}},
\author[B]{\fnms{Ryad}~\snm{Ghanam}\ead[label=e3]{raghanam@vcu.edu}}
and
\author[C,A]{\fnms{Anton}~\snm{Westveld}\ead[label=e4]{anton.westveld@anu.edu.au}}

\address[A]{Statistical Sciences and Operations Research, Virginia Commonwealth University, Richmond, Virginia, USA \printead[presep={ \ \ }]{e1,e2}}

\address[B]{Liberal Arts and Sciences, Virginia Commonwealth University School of the Arts in Qatar, Doha, Qatar \printead[presep={ \ \ }]{e3}}

\address[C]{Research School of Finance, Actuarial Studies and Statistics, Australian National University, Canberra, AUS \printead[presep={ \ \ }]{e4}}
\end{aug}

\begin{abstract}
Variable selection in linear regression models has been a problem since hypothesis testing began.  Which variables to include or exclude from a model is not an easy task.  Techniques such as Forward, Backward, Stepwise Regression sequentially add or delete variables from a model.  Penalized likelihood methods such as AIC, BIC, etc. seek to choose variables that have a significant contribution to the likelihood.  Penalized sum of square methods such as LASSO and Elastic Net have been used to penalize small coefficients to only allow variables with large coefficients in the model.  This work introduces an Artificial Intelligence approach to model selection where an ANN is trained to determine the significance of the variables based on OLS estimates.  A simulation study shows the accuracy across various sample sizes and variances.  Furthermore, a simulation study is conducted to compare the performance of the approach against Forward, Backward, AIC, BIC and LASSO.  The approach is illustrated using a dataset from the World Health Organization regarding Life Expectancy.  A github link is provided to the pretrained ANN that can handle up to 100 predictor variables, the original WHO dataset and the subset used in this work.
\end{abstract}


\end{frontmatter}

\section{Introduction}\label{sec.intro}

Selecting an appropriate subset of predictors in linear regression remains a longstanding and practically important problem in statistics. The introduction of hypothesis-testing–based regression procedures in the mid-20th century quickly revealed the difficulty of deciding which variables meaningfully contribute to a model and which do not. Although the regression framework itself is straightforward, identifying a stable and interpretable subset of predictors is often far more complicated, particularly in settings where many variables exhibit weak, correlated, or context-specific effects.  In this work the term ``significant'' is avoided as it already has well studied and understood meaning in the literature.  Instead, the term ``active'' (versus ``inactive'') will be used if a variable is selected by any of the methods presented here.

Classical variable-selection procedures such as Forward Selection, Backward Elimination, and Stepwise Regression \citep{Draper1998, Miller2002} continue to be widely used because of their simplicity and ease of implementation. These methods proceed by sequentially adding or removing predictors based on significance tests or changes in the residual sum of squares. While accessible, they are well known to be sensitive to collinearity, to the order in which variables are entered, and to the repeated use of hypothesis tests along a single path. As a result, different datasets—or even slightly perturbed versions of the same dataset—can yield different selected models.

An alternative line of work relies on information criteria such as the Akaike Information Criterion \citep{Akaike1974} and the Bayesian Information Criterion \citep{Schwarz1978}. These approaches penalize model complexity with the goal of balancing fit and parsimony. AIC generally favors models with stronger predictive performance, while BIC places heavier emphasis on selecting the true underlying model under suitable assumptions. Despite their theoretical foundations, these methods require evaluating a large number of candidate models and do not inherently address nonlinear structure or multicollinearity unless the modeler introduces such structure explicitly.

Penalized regression techniques such as the LASSO 
\citep{Tibshirani1996} and Elastic Net \citep{Zou2005} provide still another framework for variable selection by shrinking coefficient estimates toward zero. These approaches handle high-dimensional settings more gracefully and can outperform stepwise and information-criterion methods when many variables are correlated. However, they require careful tuning-parameter selection and may include or exclude variables inconsistently, especially when several predictors exert similar effects.

Recent developments in machine learning provide a new perspective on this classical statistical problem. Artificial neural networks (ANNs) have demonstrated exceptional capability in detecting nonlinear structure, capturing complex interactions, and solving classification tasks across a wide variety of domains. Their success raises a natural question: can a neural network learn to distinguish important regression predictors from noise and thus serve as a model-selection tool? While neural networks have been used extensively for prediction, their role in model selection has received comparatively little attention.

The present paper introduces a method that reframes variable selection as a classification problem. The core idea is to simulate large numbers of linear-model settings under varying sample sizes, variances, and coefficient structures, compute ordinary least squares estimates and test statistics, and train an ANN to recognize the pattern of evidence that distinguishes truly active predictors from inactive ones. This approach bypasses the need for stepwise heuristics, information-criterion comparisons across thousands of models, or penalized likelihood tuning. Once trained, the neural network outputs a binary classification for each coefficient—effectively an inclusion indicator—based solely on the estimated regression quantities. Because the ANN is trained on a diverse and carefully constructed set of simulated models, it develops a flexible decision rule that generalizes well to new data.

To evaluate the proposed approach, we conduct an extensive simulation study comparing its performance with Forward Selection, Backward Elimination, AIC, BIC, and LASSO. The method demonstrates high accuracy in identifying meaningful predictors while maintaining a relatively low false-positive rate across a wide range of scenarios. Importantly, through a padding and masking mechanism, the network can accommodate any number of predictors up to 100 without retraining. A real-data illustration using World Health Organization life-expectancy data further demonstrates the method’s interpretability and practical viability.

By integrating modern machine-learning techniques with classical regression concepts, this work aims to broaden the toolkit available for variable selection. While the ANN approach does not supplant the well-established statistical methods that have guided practice for decades, it offers a promising complementary perspective—particularly in situations where model structure is complex, predictors are highly correlated, or traditional techniques yield unstable results.

\section{Background and Related Work}\label{sec.background}

Variable selection remains a fundamental challenge in regression analysis, especially when the number of candidate predictors is large and the underlying relationships are not straightforward. In this work, each candidate model is represented by a simple indicator structure,
\[
M_k=
\begin{cases}
1, & X_k \text{ included},\\[4pt]
0, & X_k \text{ excluded},
\end{cases}
\]
which highlights the exponential model space: with $p$ predictors, there are $2^p$ possible subsets. Exhaustively evaluating all such models is rarely feasible, which is why heuristic and penalized approaches remain widely used.

Classical procedures such as forward, backward, and stepwise selection are computationally appealing, but their limitations are well documented. They rely on sequential hypothesis testing, are sensitive to correlated predictors, and often produce unstable model selections. Likelihood-based tools such as AIC \citep{Akaike1974} and BIC \citep{Schwarz1978}, built on the standard linear regression model
\[
Y_i = \beta_0 + \beta_1 X_{i,1} + \cdots + \beta_p X_{i,p} + \varepsilon_i,  \ \ \epsilon_i \iid \textrm{normal} (0,\sigma^2), 
\]
attempt to provide more principled model comparisons by penalizing model complexity. However, even these criteria must be computed across many candidate models and may favor conflicting subsets under different sample sizes and noise conditions.

Penalized regression methods, particularly the LASSO \citep{Tibshirani1996}, introduce sparsity directly by minimizing
\[
\min_{\beta}
\left\{
\text{RSS} + \lambda \sum_{j=1}^p |\beta_j|
\right\},
\]
where RSS $=\frac{1}{n} \sum_{i=1}^n (y_i - \hat{y}_i)^2$. The sparsity encourages smaller models and handles correlated predictors more gracefully. Nonetheless, the LASSO depends on tuning parameters and may leave small nonzero coefficients in the model, making it less suitable for settings where clear binary inclusion decisions are needed.

The introduction of this work includes simulation-based illustrations that clearly demonstrate the limitations of fixed analytic rules. More importantly, Figure \ref{fig:ANNMap} below shows how strongly the $t$-statistics for active and inactive predictors can overlap across different simulated settings. These overlaps explain why classical threshold-based rules---such as selecting predictors with large t-values---often misclassify noise as genuine signal or overlook weak but real effects. These observations motivate the need for a flexible, data-driven approach capable of adapting to different noise levels, signal strengths, and correlation structures. 

\begin{figure}[htbp]
\centering
\includegraphics[width=0.9\textwidth]{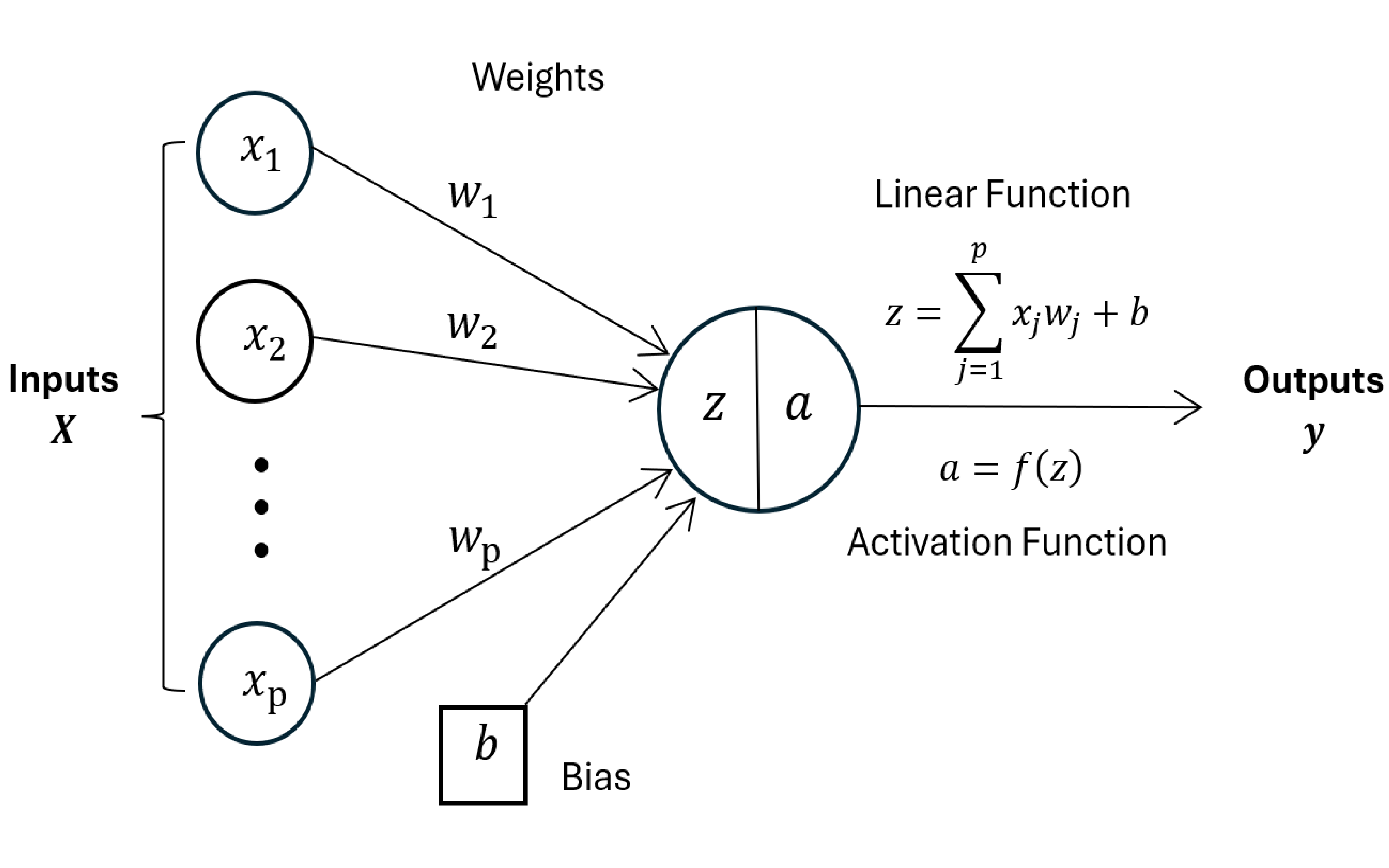}  
\caption{The basic architecture of an artificial neuron}
\label{fig:ANNneuron}
\end{figure}

\section{Neural Networks}\label{sec.nueralnet}

Neural networks offer a versatile framework for learning patterns in data without relying on fixed analytic thresholds. In this work, simulated datasets generated from the linear model
\[
Y_i = \beta_0 + \beta_1 X_{i,1} + \cdots + \beta_p X_{i,p} + \varepsilon_i,  \ \ \epsilon_i \iid \textrm{normal} (0,\sigma^2), 
\]
produce $t$-statistics that serve as the main inputs to the classifier:
\[
t_j = \frac{\widehat{\beta}_j}{\mathrm{SE}(\widehat{\beta}_j)}.
\]

These $t$-values capture both effect size and sampling variability and are widely used in traditional variable-selection criteria. However, as illustrated earlier in Figure \ref{fig:ANNMap}, the distributions of $t$-values for active and inactive predictors often overlap---sometimes substantially---making fixed cutoffs unreliable across diverse scenarios.

To address this, the study employs a neural network (Figure \ref{fig:ANNneuron}) that transforms these $t$-values using a sigmoid activation function
\[
\sigma(z)=\frac{1}{1+e^{-z}}.
\]

Training the neural network requires determining how each weight influences the overall loss. This is carried out using backpropagation, where the gradient of the loss with respect to a weight is computed via the chain rule:
\[
\frac{\partial L}{\partial w}
=
\frac{\partial L}{\partial \hat{y}}
\cdot
\frac{\partial \hat{y}}{\partial z}
\cdot
\frac{\partial z}{\partial w},
\]
and the corresponding weight update rule,
\[
w \leftarrow w - \eta\,\frac{\partial L}{\partial w},
\]
with $\eta$ denoting the learning rate.

Because the network is trained on simulated data covering a wide range of sample sizes, noise levels, and effect patterns, it learns how $t$-statistics behave under conditions where classical methods perform poorly. By exposing the model to diverse scenarios---including those where active and inactive $t$-values overlap---the network gains the ability to adapt its classification rule to the underlying structure of the data.
 
This combination of simulation-driven learning and neural-network methodology provides a flexible alternative to traditional variable-selection tools and forms the basis for the approach developed in the remainder of the work.

Neural network models are often built using software tools that simplify training and testing.  TensorFlow \citep{Abadi2016} is an open-source tool used to perform large mathematical calculations and train neural networks.  It can run on different types of hardware, such as CPUs and GPUs.  Keras \citep{Chollet2021} works on top of TensorFlow and makes neural networks easier to build and train by using simple and organized layers.  Because of its simplicity, TensorFlow with Keras is widely used in research and practice.  PyTorch \citep{Paszke2019} is another common deep learning tool that allows models to be built step by step while the program is running, unlike Keras within TensorFlow, which makes models easier to debug.  PyTorch also runs efficiently on GPUs and is widely used by researchers. In this work, Keras version 3.11.2 within TensorFlow version 2.18.1 was used rather than building the AI model step-by-step from scratch.

This work is organized in the following manner. Section~\ref{sec.aims} explains the proposed method for conducting model selection using ANNs. Section~\ref{sec.sim1} shows a simulation study of the Type I and Type II error rates along with a power study of the proposed method, LASSO, AIC, BIC, Forward and Backward selection methods. Section~\ref{sec.scalability} discusses scalability. Section~\ref{sec.who} applies the methods to a real world dataset from the World Health Organization regarding life expectancy. Finally, Section~\ref{sec.conclusion} provides a discussion, conclusion and future work.

\section{AI Model Selection}\label{sec.aims}
All model selection algorithms have the objective of finding the $X$ variables that are related to $Y$.  Recent developments in ANN packages have made this tool widely available to researchers.  Packages such as keras in Tensorflow and PyTorch allow for one to construct and train AI models using the Python programming language.  The HuggingFace website also has a large repository of tools for implementing AI.  Model selection can be viewed as a $p$ dimension classification problem where each predictor variable is classified as $0$ or $1$.  Current ANN technology can easily handle this task. 

\begin{figure}[ht]
    \begin{center}
  \includegraphics[width=0.9\textwidth]{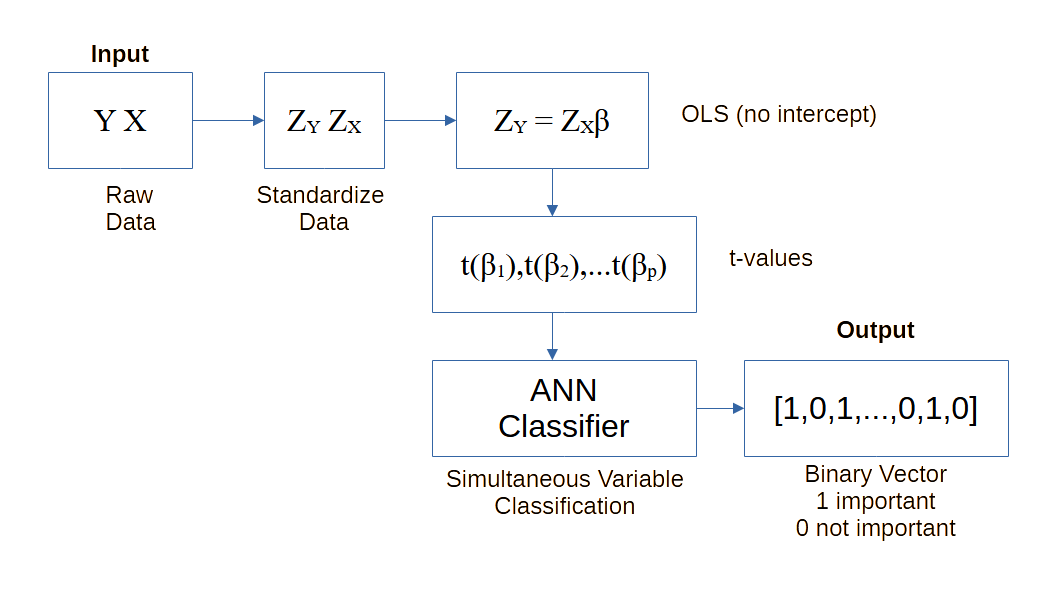}  
  \caption{Flowchart of model selection algorithm.}
\label{fig:ANNMap}
  \end{center}
\end{figure}

Figure~\ref{fig:ANNMap} shows a flowchart of the algorithm for ANN models selection.  Starting with the raw data each variable is standardized using a z-score resulting in $Z_{X_1}, Z_{X_2}, ... Z_{X_p}$ for the $p$ predictors and $Z_Y$ for the response.  Next a standardized regression with out intercept is calculated.  Note that an intercept is not needed as $E[Z_Y]=0$ when all predictors are zero.  From here, obtain the vector t-values for each of the estimated regression coefficients $(t\{ \hat{\beta}_1 \}, t\{ \hat{\beta}_2 \},...,t\{ \hat{\beta}_p \})$.  This vector is then fed into the ANN simultaneously resulting in a binary vector of length $p$ where a value of one corresponds to an important variable and zero corresponds to not important.

\subsection{ANN Training}
The key to this approach is training an ANN to be able to detect important variables.  Here a synthetic data approach is used to generate thousands of datasets which have some variables as active and some inactive.  For clarity, The term ``dataset`` will be used to refer to a single synthetic regression dataset, the term ``training set'' corresponds to the set of all synthetic regression datasets used for training.  Similarly, the term ``validation set'' corresponds to all synthetic regression datasets used for validation.  To conduct our training dataset it is imperative that the important variables are known as these are our the outcome variable of the ANN.  Hence in generating the datasets the important variables are decided before the regression coefficients are generated.  The following algorithm is used to generate both the training and validation sets.
\begin{enumerate}
    \item Generate a true binary indicator vector for this dataset:  $\Gamma_k$ where $\Gamma_{kj}$ is 0 or 1 with equal probability.  (This is the output for one instance of the ANN).
    \item Generate the sample size $n_k \sim \text{DiscreteUniform}(20,2000)$.
    \item Generate $p$ predictors from $Z_{ij} \sim N(0,1),~j=1,2,...p,~i=1,2,...,n$ to form regression matrix $\mathbf{X}$.
    \item For each $\Gamma_{kj}=1$ generate a true regression coefficient $\beta_j$ such that it is not near 0 using $\beta_k^{(c)} \sim Uniform( (-5,-0.01) \cup (0.01,5) )$.  If $\Gamma_{kj}=0$, then $\beta_k = 0$.
    \item Create the $Y = \mathbf{X}\beta + \epsilon$ where $\epsilon_i \iid N(0, \sigma^2)$.  
    \item For the $Y$ and $X$ dataset fit the ordinary least squares regression and without an intercept to obtain the vector of the t-values of the estimated regression coefficients:  $T_k = $ $\left(t\{ \hat{\beta}_1 \}, t\{ \hat{\beta}_2 \},...,t\{ \hat{\beta}_p \} \right)$.  (This is the input for one instance of the ANN)
    \item Repeat the above steps and generate 10,000 datasets to create the training set input $\mathbf{T}^{t}  =\{T_1, T_2,...,T_{N}\}$ and output $\mathbf{\Gamma}^{t} = \{\Gamma_1, \Gamma_2,...,\Gamma_{N}\}$, and generate 1000 validation datasets to create validation set input $\mathbf{T}^{V}  =\{T_1, T_2,...,T_{N}\}$ and output $\mathbf{\Gamma}^{v} = \{\Gamma_1, \Gamma_2,...,\Gamma_{N}\}$
    \item Train an ANN with  as the input $\mathbf{T}^{t}$ and $\mathbf{\Gamma}^{t}$ as the output.  
    \item Pass the validation datasets through the trained ANN to obtain $\hat{\Gamma}^{v}$.  From this we will get correct positive (CP), correct negative (CN), false positive (FP) and false negative (FN) rates for the classifications. 
\end{enumerate}

\section{Simulation Study}\label{sec.sim1}

To assess the viability of the approach a training set is created using $p=10$ and containing $N=10,000$ synthetic datasets.  Similarly, 1,000 simulated validation datasets were created for this setting. Using the training set an ANN with one layer with 10 sigmoid nodes which are fully connected to both the input layer and the output layer is trained via backpropagation with 10,000 epochs. The ANN model is built and trained using the Keras and TensorFlow framework.  Dataset sample sizes $n=50, 250$, and $1,000$ and variances $\sigma^2=0.01, 0.1,$ and $0.5$ are considered to determine the effect of sample size and model error variance on the accuracy rates, specifically, Correct Positive (CP), Correct Negative (CN), False Positive (FP) and False Negative (FN).  For comparison, Lasso, stepwise forward, stepwise backward, AIC, and BIC are also performed on the same datasets along with their corresponding accuracy rates.

\begin{table}[hbt!]
\scriptsize   
\setlength{\tabcolsep}{4pt}   
\renewcommand{\arraystretch}{0.6} 
\begin{threeparttable}
\caption{Confusion matrix results (correct positives (CP), correct negatives (CN), false positives (FP), and false negatives (FN)) computed from 1000 simulated validation datasets. Results are recorded for varying sample sizes $(n=50, 250, 1000)$ and noise levels $(\sigma^2=0.01, 0.1, 0.5)$. The table shows how ANN, Lasso, stepwise forward, stepwise backward, AIC, and BIC perform in detecting overall effect sizes under these conditions. Based on 1,000 simulated validation datasets.}
\label{tab:cm_results_all}
\begin{tabular}{l*{6}{l}}
\toprule
\textbf{Size $n$} & \textbf{variance $\sigma^2$} & \textbf{Method} & \textbf{CP} & \textbf{CN} & \textbf{FP} & \textbf{FN} \\
\midrule
50 & 0.01 & ANN & 0.920 & 1 & 0 & 0.08 \\
& & Lasso            & 0.928 & 0.962 & 0.038 & 0.072 \\
& & Stepwise forward & 1 & 0.948 & 0.052 & 0 \\
& & Stepwise backward & 1 & 0.948 & 0.052 & 0 \\
& & AIC              & 1 & 0.793 & 0.207 & 0 \\
& & BIC              & 1 & 0.924 & 0.076 & 0 \\
\midrule
50 & 0.1 & ANN & 0.755 & 0.996 & 0.004 & 0.245 \\
& & Lasso            & 0.929 & 0.552 & 0.445 & 0.071 \\
& & Stepwise forward & 0.813 & 0.95 & 0.05 & 0.187 \\
& & Stepwise backward & 0.816 & 0.943 & 0.057 & 0.184 \\
& & AIC              & 0.881 & 0.788 & 0.212 & 0.119 \\
& & BIC              & 0.829 & 0.927 & 0.073 & 0.171 \\
\midrule
50 & 0.5 & ANN & 0.391 & 0.989 & 0.011 & 0.609 \\
& & Lasso            & 0.949 & 0.118 & 0.882 & 0.071 \\
& & Stepwise forward & 0.496 & 0.949 & 0.051 & 0.504 \\
& & Stepwise backward & 0.513 & 0.94 & 0.06 & 0.487 \\
& & AIC              & 0.65 & 0.789 & 0.211 & 0.35 \\
& & BIC              & 0.528 & 0.934 & 0.066 & 0.472 \\
\midrule
250 & 0.01 & ANN & 0.995 & 1 & 0 & 0.005 \\
& & Lasso            & 0.937 & 1 & 0 & 0.063 \\
& & Stepwise forward & 1 & 0.938 & 0.062 & 0 \\
& & Stepwise backward & 1 & 0.938 & 0.062 & 0 \\
& & AIC              & 1 & 0.828 & 0.172 & 0 \\
& & BIC              & 1 & 0.978 & 0.022 & 0 \\
\midrule
250 & 0.1 & ANN & 0.867 & 0.997 & 0.003 & 0.133 \\
& & Lasso            & 0.938 & 0.854 & 0.146 & 0.062 \\
& & Stepwise forward & 0.921 & 0.941 & 0.059 & 0.079 \\
& & Stepwise backward & 0.921 & 0.941 & 0.059 & 0.079 \\
& & AIC              & 0.952 & 0.817 & 0.183 & 0.048 \\
& & BIC              & 0.905 & 0.972 & 0.028 & 0.095 \\
\midrule
250 & 0.5 & ANN & 0.64 & 0.995 & 0.005 & 0.359 \\
& & Lasso            & 0.941 & 0.232 & 0.768 & 0.059 \\
& & Stepwise forward & 0.711 & 0.944 & 0.056 & 0.289 \\
& & Stepwise backward & 0.711 & 0.944 & 0.056 & 0.288 \\
& & AIC              & 0.778 & 0.84 & 0.16 & 0.222 \\
& & BIC              & 0.677 & 0.979 & 0.021 & 0.326 \\
\midrule
1000 & 0.01 & ANN & 1 & 1 & 0 & 0 \\
& & Lasso            & 0.94 & 1 & 0 & 0.06 \\
& & Stepwise forward & 1 & 0.947 & 0.053 & 0 \\
& & Stepwise backward & 1 & 0.947 & 0.053 & 0 \\
& & AIC              & 1 & 0.83 & 0.17 & 0 \\
& & BIC              & 1 & 0.987 & 0.013 & 0 \\
\midrule
1000 & 0.1 & ANN & 0.896 & 1 & 0 & 0.104 \\
& & Lasso            & 0.945 & 0.996 & 0.004 & 0.054 \\
& & Stepwise forward & 0.986 & 0.944 & 0.056 & 0.014 \\
& & Stepwise backward & 0.986 & 0.944 & 0.056 & 0.014 \\
& & AIC              & 0.996 & 0.836 & 0.164 & 0.004 \\
& & BIC              & 0.969 & 0.986 & 0.14 & 0.031 \\
\midrule
1000 & 0.5 & ANN & 0.75 & 0.998 & 0.002 & 0.25 \\
& & Lasso            & 0.937 & 0.462 & 0.538 & 0.063 \\
& & Stepwise forward & 0.813 & 0.951 & 0.049 & 0.187 \\
& & Stepwise backward & 0.813 & 0.951 & 0.049 & 0.187 \\
& & AIC              & 0.861 & 0.839 & 0.161 & 0.139 \\
& & BIC              & 0.763 & 0.994 & 0.006 & 0.237 \\
\bottomrule
\end{tabular}
\end{threeparttable}
\end{table}

Table \ref{tab:cm_results_all} shows the classification rates of our ANN model along with Lasso, Stepwise Forward, Stepwise Backward, AIC, and BIC on validation data created using the same algorithm as the training data.  To ensure a fair comparison Lasso the best $\lambda$ value was determined via cross-validation. The samples sizes $n=50, 250$ and $1,000$ and variances $\sigma^2=0.01,0.1$ and $0.5$ were examined to determine the Correct Positive (CP), Correct Negative (CN), False Positives (FP) and False Negatives (FN) for each of the methods.  All values in the table are based on 1,000 simulated datasets under each condition.  Notice that when the variance increased, the ANN approach has a lower CP rate.  Specifically, when $n=50$ with $\sigma^2 = 0.5$ the ANN has a CP rate of 0.391, for $n=250$ and $\sigma^2 =0.5$ the ANN has a CP rate of 0.64 and for $n=1,000$ and $\sigma^2=0.5$ the CP rate is 0.75.  This is lower than all of the other methods considered here. In contrast when $\sigma^2$ is low the CP rate is high with the lowest CP rate of 0.92. However, where the ANN excels above all other methods is the CN classification with the lowest CN rate of 0.989 across all sample sizes and variance settings. This translates directly to a extremely low FP rate as well.

\begin{figure}[htbp]
\centering
\includegraphics[width=0.9\linewidth]{{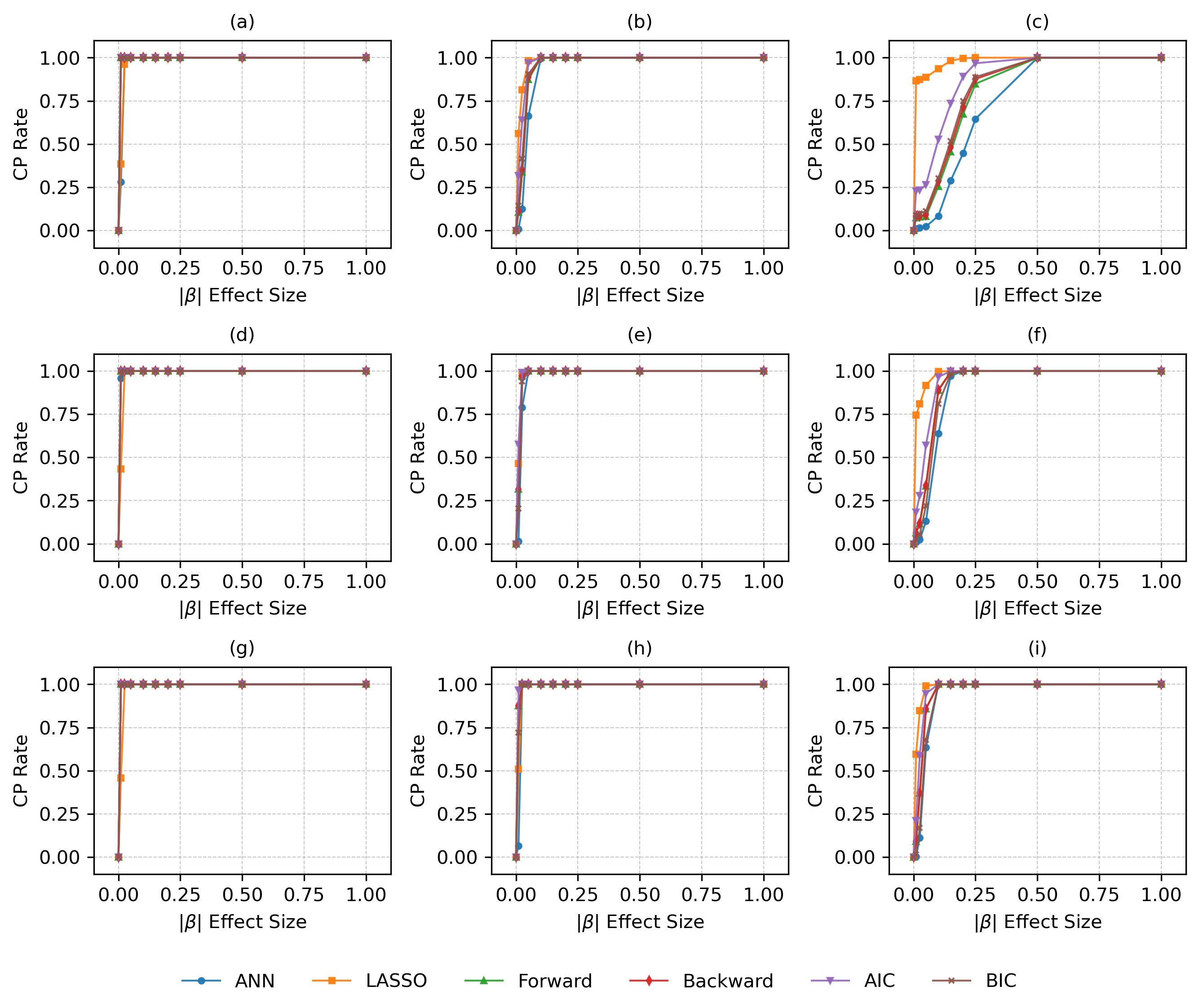}}
\caption{Correct Positive Rates (CP) against Effect Sizes $|\beta|$ across different levels of sample sizes $n$ and variances $\sigma^2$. Each subplot corresponds to a unique combination of ($n, \sigma^2$) such as: (a) (50,0.1), (b) (50,0.1), (c) (50,0.5), (d) (250,0.01), (e) (250,0.1), (f) (250,0.5), (g) (1000,0.01), (h) (1000,0.1), and (i) (1000,0.5). The curves show the power of well known selection methods in selecting the significant variables. Six methods are compared based on OLS estimates: ANN, Lasso, Stepwise Forward, Stepwise Backward, AIC, and BIC.  Based on 1,000 simulation for each $(n,\sigma)$ combination.}
\label{fig:cp_rates_all}
\end{figure}

Using the trained ANN model, a power study is performed in order to assess the ability of the method to detect various effect levels.  Again, the datasets are generated with levels of $(n = 50, 250, 1000$) and $(\sigma^2 = 0.01, 0.1, 0.5)$.  For the generation of $Y$, a vector of true regression coefficients is specified as $\beta = [1, 0.5, 0.25, 0.2, 0.15, 0.1, 0.05, 0.025, 0.01, 0]$, which reflect a reasonable range of effect sizes.  Note that the associated covariates are standardized and hence the first coefficient corresponds to a one standard deviation change in $y$ when there is a one standard deviation change in $x_1$.  The last value in the in the vector is $0$ to have a reference to when a variable is not significant.  The results are based on 1,000 simulated validation datasets for each $n$ and $\sigma^2$ combination.  The ANN, LASSO, Forward, Backward, AIC and BIC methods are run on each of the datasets and the proportion of the simulated datasets where the coefficient under consideration is deemed important is recorded and plotted together.  The results of this are found in Figure \ref{fig:cp_rates_all}.  In panel (a) $n=50$ and $\sigma^2=0.1$ and shows that all methods are competitive with each method being able to detect very small effect sizes at similar rates.  In the figure $\sigma^2$ increases across the panels to the right and the sample size increases downwards across the panels.  Panels (c,f,i) show that for high variance $\sigma^2 = 0.5$ there is a difference in performance of the methods.  Especially in panel (c) where LASSO performs the best as it has the highest CP rate across all the coefficient levels.  And similarly, ANN performs the worst as it lowest across all effect levels.  However, as the sample size increases in panels (f) and (i) the differences between the methods becomes minimal.  Hence for cases when there is low variance all methods perform similarly.

As computation time is often a concern using selection methods, especially for large datasets, the computation times were recorded for the power study.  These were averaged across all simulations at each sample size for each method.  Table \ref{tab:compute_times} shows the compute times in seconds.  Notice that among all methods LASSO performed best at less than 0.00092 seconds for all sample sizes. Next is stepwise backwards with times less than 0.0066 seconds.  Followed by the ANN approach with times less than 0.0679 seconds.  Stepwise forward is has computation times less than 0.1558.  Both AIC and BIC have larger computational times as they need to compute more models (specifically $(2^p)$ models) than any of the other methods with times less than 1.068 seconds.  It should be noted that the LASSO method will require additional calculations as once the correct variables are selected the model containing those variables sill need to be calculated.

\begin{table}[hbt!]
\scriptsize   
\setlength{\tabcolsep}{2pt}   
\renewcommand{\arraystretch}{0.6} 
\begin{threeparttable}
\caption{Average computational times for the simulated validation datasets in seconds are presented sample sizes $(n=50, 250, 1000)$ for each of the selection methods ANN, Lasso, stepwise forward, stepwise backward, AIC, and BIC.  Results based on 3,000 simulations for each sample size.}
\label{tab:compute_times}
\begin{tabular}{l*{6}{l}}
\toprule
\textbf{$n$} & \textbf{ANN} & \textbf{Lasso} & \textbf{Stepwise forward} & \textbf{Stepwise backward} & \textbf{AIC} & \textbf{BIC} \\
\midrule
50 & 0.0699 & 0.00089 & 0.1059 & 0.0050 & 0.2321 & 0.2321 \\
250 & 0.0664 & 0.00089 & 0.1183 & 0.0057 & 0.3891 & 0.3891 \\
1000 & 0.0679 & 0.00092 & 0.1558 & 0.0066 & 1.0680 & 1.0680 \\
\bottomrule
\end{tabular}
\end{threeparttable}
\end{table}    

\section{Scalability}\label{sec.scalability}
\begin{figure}[ht]
    \begin{center}
  \includegraphics[width=0.9\textwidth]{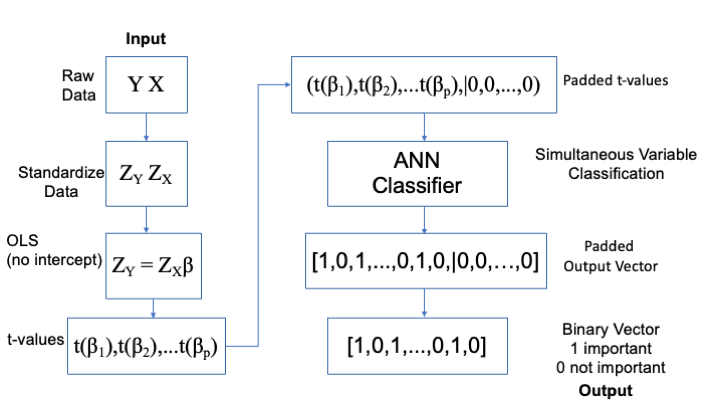}  
  \caption{Flowchart of padded model selection algorithm.}
\label{fig:ANNMapPadded}
  \end{center}
\end{figure}

The issue of scalability needs to be addressed as this method needs to be applied to a wide range of possible scenarios.  In the previous section the ANN approach was explored with the number of covariates, $p=10$, for simplicity.  However, the approach is not useful if one must have exactly $10$ covariates to use the method.  To extend the approach, the method of ``padding'' \citep{Dwarampudi2019} is used to build an ANN model that can accept up to 100 covariates.  Suppose $t(\beta_1),t(\beta_2),...,t(\beta_p)$ is the vector of t-values associated with the full model using all $p$ covariates (here $p<100$).  To ``pad'' the t-values, one concatenates a vector of zeros of length $100-p$ to the t-value vector.  This becomes the input vector to the ANN.  Similarly, once the output of the ANN is produced the vector of important variables will be of length 100 in which the first $p$ entries are used to determine the important variables in the original dataset. Figure~\ref{fig:ANNMapPadded} shows the algorithm to determine the important variables.  It is similar to the algorithm in Figure~\ref{fig:ANNMap} except that the t-values are padded after they are calculated via OLS and the output is from a unpadded binary vector from the ANN classifier.  The algorithm below provides the specifics of the training and validation sets and training and validation methods. 

To train the ANN in this case a similar approach was used as in the previous ANN, however, there must be many more models in both our training set and our validation sets.


\begin{enumerate}
    \item Generate a true binary indicator vector for this dataset:  $\Gamma_k$ where $\Gamma_{kj}$ is 0 or 1 with equal probability. 
    \item Generate the sample size $n_k \sim \text{DiscreteUniform}(20,2000)$.
    \item Generate $p$ predictors from $Z_{ij} \sim N(0,1),~j=1,2,...,p,~i=1,2,...,n$ to form regression matrix $\mathbf{X}$, where $p=[100,99,...,1]$.
    \item For each $\Gamma_{kj}=1$ generate a true regression coefficient $\beta_j$ such that it is not near 0 using $\beta_k^{(c)} \sim Uniform( (-5,-0.01) \cup (0.01,5) )$.  If $\Gamma_{kj}=0$, then $\beta_k = 0$.
    \item Create the $Y = \mathbf{X}\beta + \epsilon$ where $\epsilon_i \iid N(0, \sigma^2)$.  
    \item For the $Y$ and $X$ dataset fit the ordinary least squares regression and without an intercept to obtain the vector of the t-values of the estimated regression coefficients:  $T_k = $ $\left(t\{ \hat{\beta}_1 \}, t\{ \hat{\beta}_2 \},...,t\{ \hat{\beta}_p \} \right)$.
    \item Pad the vector of the t-values with a vector of zeros of length $100-p$ such as $T_k = $ $\left(t\{ \hat{\beta}_1 \}, t\{ \hat{\beta}_2 \},...,t\{ \hat{\beta}_p \} | \mathbf{0}_{100-p} \right)$. (This is the input for one instance of the ANN)
    \item Pad the true binary indicator vector with a vector of zeros of length $100-p$ such as $\Gamma_k = $ $\left( \Gamma_1, \Gamma_2,...,\Gamma_p | \mathbf{0}_{100-p} \right)$. (This is the output for one instance of the ANN). 
    \item Repeat the above steps and generate 100,000 datasets to create the training set input $\mathbf{T}^{t}  =\{T_1, T_2,...,T_{N}\}$ and output $\mathbf{\Gamma}^{t} = \{\Gamma_1, \Gamma_2,...,\Gamma_{N}\}$ and validation set input $\mathbf{T}^{V}  =\{T_1, T_2,...,T_{N}\}$ and output $\mathbf{\Gamma}^{v} = \{\Gamma_1, \Gamma_2,...,\Gamma_{N}\}$
    \item Train an ANN with as the input $\mathbf{T}^{t}$ and $\mathbf{\Gamma}^{t}$ as the output.  
    \item Pass the validation datasets through the trained ANN to obtain $\hat{\Gamma}^{v}$.  From this we will get correct positive (CP), correct negative (CN), false positive (FP) and false negative (FN) rates for the classifications. 
\end{enumerate}

The results of the training for the larger ANN is given as a confusion matrix in Table~\ref{tab:cm_ANN}.  Notice that the CP rate is 0.9998 and the CN rate is 0.9877 which indicates extremely good performance.  The FP and FN rates are very low in this model which suggests the ANN model is more than sufficient for determining which variables are important across a wide variety datasets that have continuous covariates and continuous response.

\begin{table}[hbt!]
\scriptsize   
\setlength{\tabcolsep}{2pt}   
\renewcommand{\arraystretch}{0.7} 
\begin{threeparttable}
\caption{Confusion matrix of the validation sets for the padded ANN model predictions. Values are based on $100,000$ simulated datasets.}
\label{tab:cm_ANN}
\begin{tabular}{l*{2}{l}}
\toprule
& \textbf{Actual Negative} & \textbf{Actual Positive} \\
\midrule
\textbf{Predicted Negative} & 0.9998 & 0.0123 \\
\textbf{Predicted Positive} & 0.0002 & 0.9877 \\
\bottomrule
\end{tabular}
\end{threeparttable}
\end{table}

\section{WHO Dataset}\label{sec.who}

A World Health Organization (WHO) data set is used to illustrate the proposed method performs on  real-world data. The dataset consists of several health variables for 141 countries from 2000 to 2015.  For this example, the target variable is Life Expectancy (The average number of years that a newborn can expect to live). The selected predictor variables are Adult Mortality (The probability of dying at age 15-60 years per 1000 population), Infant Deaths (The number of infant deaths per 1000 live birth), Measles (The number of Measles disease cases reported per 1000 population), BMI (The average number of Body Mass Indexes (BMI) for the adult population), Polio (The percentage of coverage among 1-year-old children who received the Polio vaccine), Diphtheria (The percentage of coverage among 1-year-old children who received diphtheria, pertussis, and tetanus vaccines (DPT3)), HIV/AIDS (The number of deaths at age 0-4 years per 1000 live births), Thinness 1-19 years (The percentage of children and adults aged 1-19 years who are underweight for their age and height), Income Composition of Resources (ICR) index (a score 0–1 represents the Human Development Index (HDI) based on income, education, and life expectancy), and Population size in millions.

\begin{table}[hbt!]
\scriptsize
\setlength{\tabcolsep}{1pt}
\renewcommand{\arraystretch}{0.7}
\begin{threeparttable}
\caption{Descriptive statistics for the selected subsets of the 2000-year and 2015-year Life Expectancy datasets with sample size $n=141$.}
\label{tab:data_summary}
\begin{tabular}{l r*{9}{r}}
\toprule
& \multicolumn{5}{c}{\textbf{2000}} & \multicolumn{5}{c}{\textbf{2015}} \\
\textbf{Variable} &
 \textbf{Mean} & \textbf{std} & \textbf{Min} & \textbf{50\%} & \textbf{Max} &
 \textbf{Mean} & \textbf{std} & \textbf{Min} & \textbf{50\%} & \textbf{Max} \\
\midrule
Life Expectancy
& 67.61 & 10.16 & 39 & 71.40 & 81.10
& 72.43 & 7.97 & 51 & 74.50 & 88 \\
Adult Mortality
& 173.23 & 143.31 & 2 & 145 & 665
& 150.68 & 95.62 & 1 & 133 & 484 \\
Infant Deaths
& 37.15 & 161.22 & 0 & 4 & 1800
& 21.56 & 83.79 & 0 & 2 & 910 \\
Measles
& 3717.47 & 9188.64 & 0 & 59 & 71093
& 1582.33 & 8736.12 & 0 & 17 & 90387 \\
BMI
& 35.26 & 19.02 & 1.40 & 41.60 & 65.60
& 43.39 & 20.94 & 3.80 & 52.60 & 75.20 \\
Polio
& 79.38 & 24.97 & 3 & 88 & 99
& 83.70 & 24.92 & 5 & 93 & 99 \\
Diphtheria
& 76.18 & 27.86 & 7 & 87 & 99
& 85 & 22.84 & 6 & 94 & 99 \\
HIV/AIDS
& 2.88 & 7.80 & 0.10 & 0.10 & 46.40
& 0.65 & 1.41 & 0.10 & 0.10 & 9.30 \\
Population (millions)
& 14.98 & 27.71 & 0.08 & 5.34 & 175.29
& 18.80 & 38.25 & 0.12 & 6.31 & 259.62 \\
Thinness 1-19 years
& 5.01 & 4.82 & 0.10 & 2.90 & 27.70
& 4.50 & 4.32 & 0.10 & 3.30 & 26.70 \\
ICR
& 0.63 & 0.17 & 0.25 & 0.65 & 0.91
& 0.71 & 0.15 & 0.35 & 0.73 & 0.95 \\
\bottomrule
\end{tabular}
\end{threeparttable}
\end{table}

\begin{figure}[htbp]
\centering
\includegraphics[width=0.7\textwidth]{{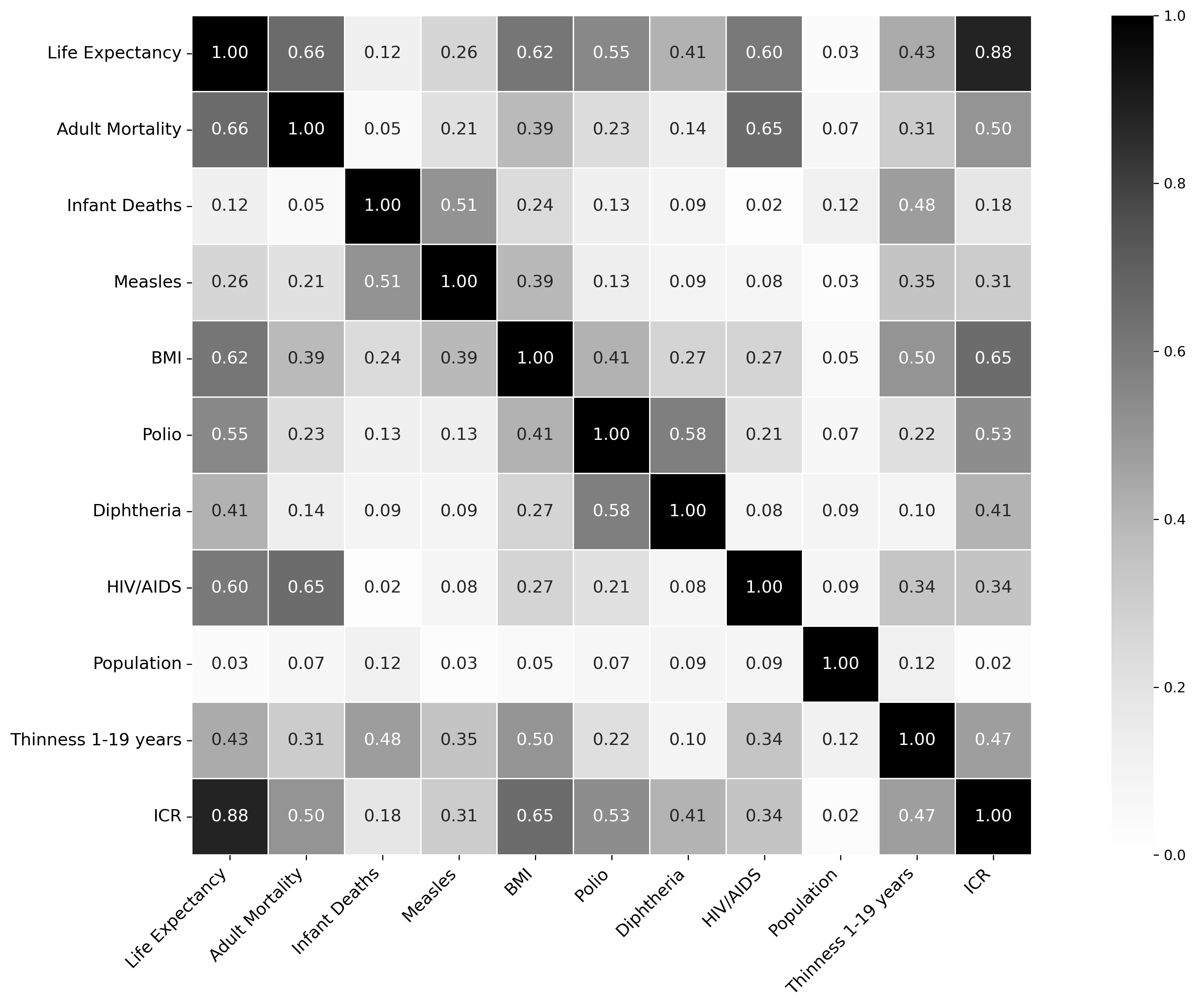}}
\caption{Pearson correlation matrix for the year 2000 dataset ($n=141$).}
\label{fig:fixed_pearson_corr_data1}
\end{figure}

\begin{figure}[htbp]
\centering
\includegraphics[width=0.7\textwidth]{{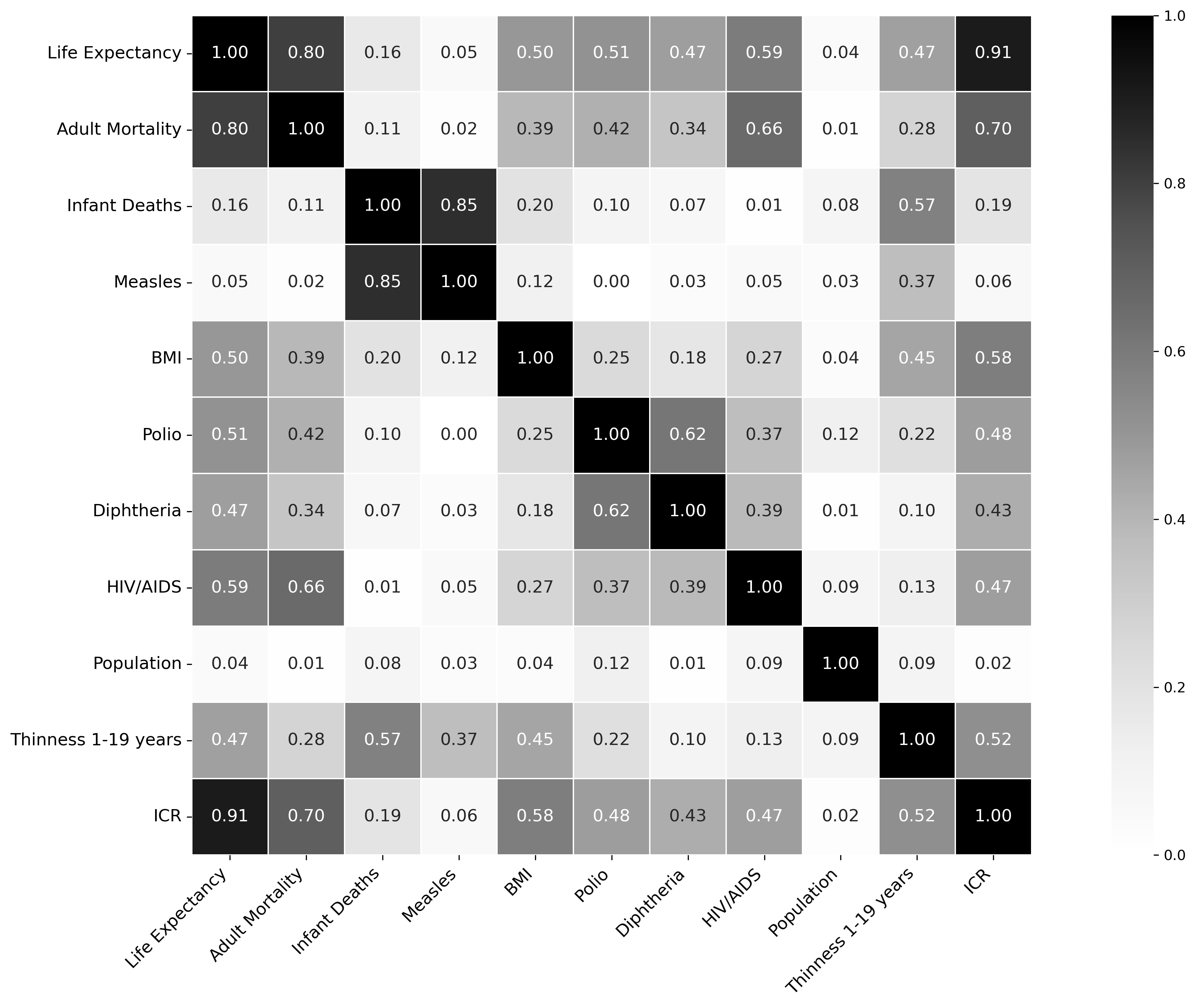}}
\caption{Pearson correlation matrix for the year 2015 dataset ($n=141$) }
\label{fig:fixed_pearson_corr_data2}
\end{figure}

Figures~\ref{fig:fixed_pearson_corr_data1} and \ref{fig:fixed_pearson_corr_data2} show the Pearson's correlation matrix of the selected subset of variables for the years 2000 and 2015, respectively.  As Infant Deaths and Under-Five Deaths have a Pearsons correlation of 0.99; and Thinness 1-19 and Thinness 5-19 have a Pearsons correlation of 0.94; lastly, ICR and Schooling has a Pearsons correlation of 0.93 several variables need to be eliminated. Among each two correlated variables, the weaker correlated variable with the output (Life Expectancy) are Under-Five Deaths, Thinness 5-9 years, and Schooling and are eliminated from the dataset.  Since the variables Infant Deaths, Measles, Polio, Diptheria, HIV/AIDS, Population and Thinness 1-19 years are highly right skewed a $\log( x + 1 )$ is used to reduce the influence of these variables.

\begin{figure}[htbp]
\centering
\includegraphics[width=0.7\textwidth]{{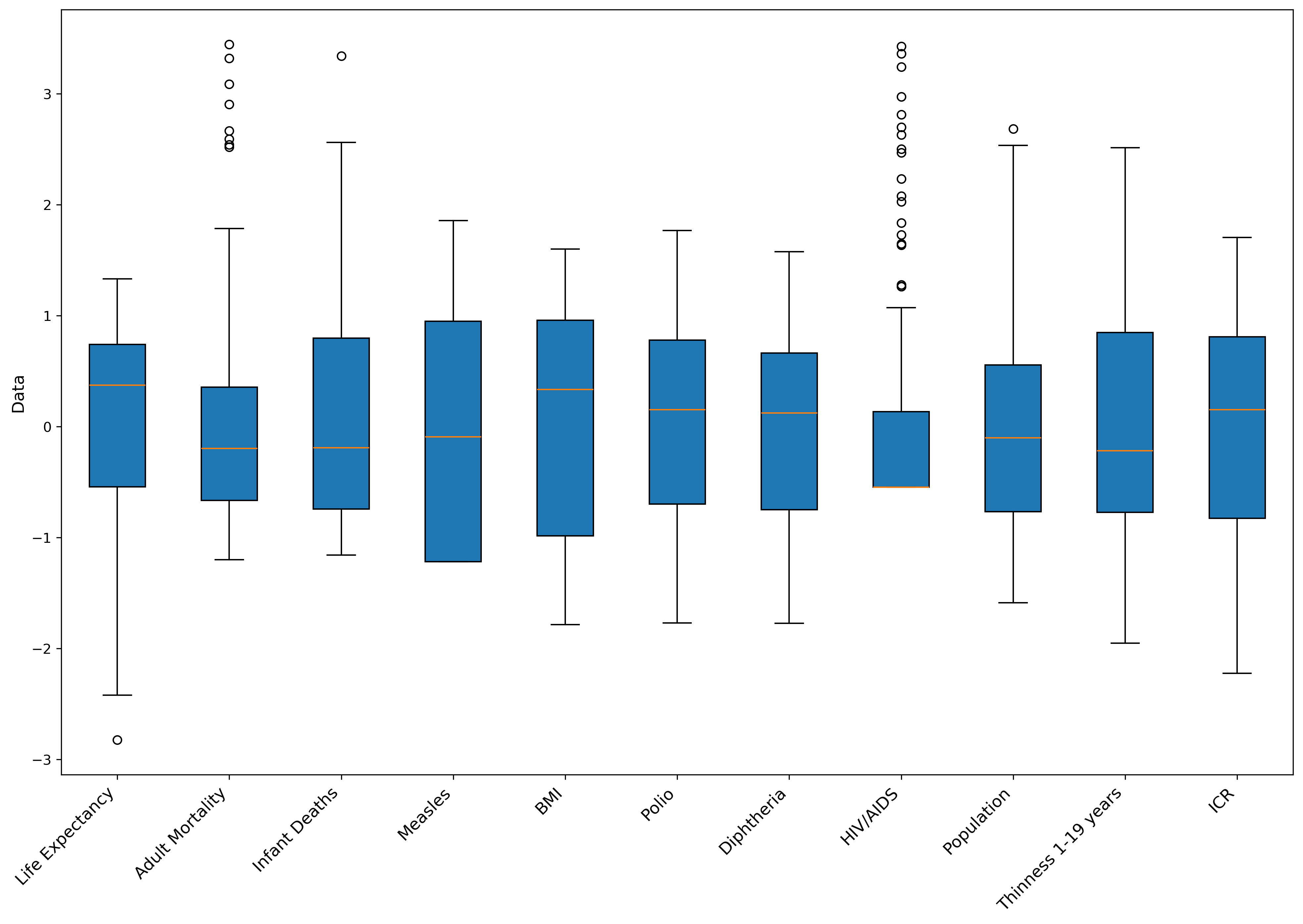}}
\caption{Box-plots of the standardized 2000-year dataset with sample size $n=141$.}
\label{fig:outlier_trans_stand_data1}
\end{figure}

\begin{figure}[htbp]
\centering
\includegraphics[width=0.7\textwidth]{{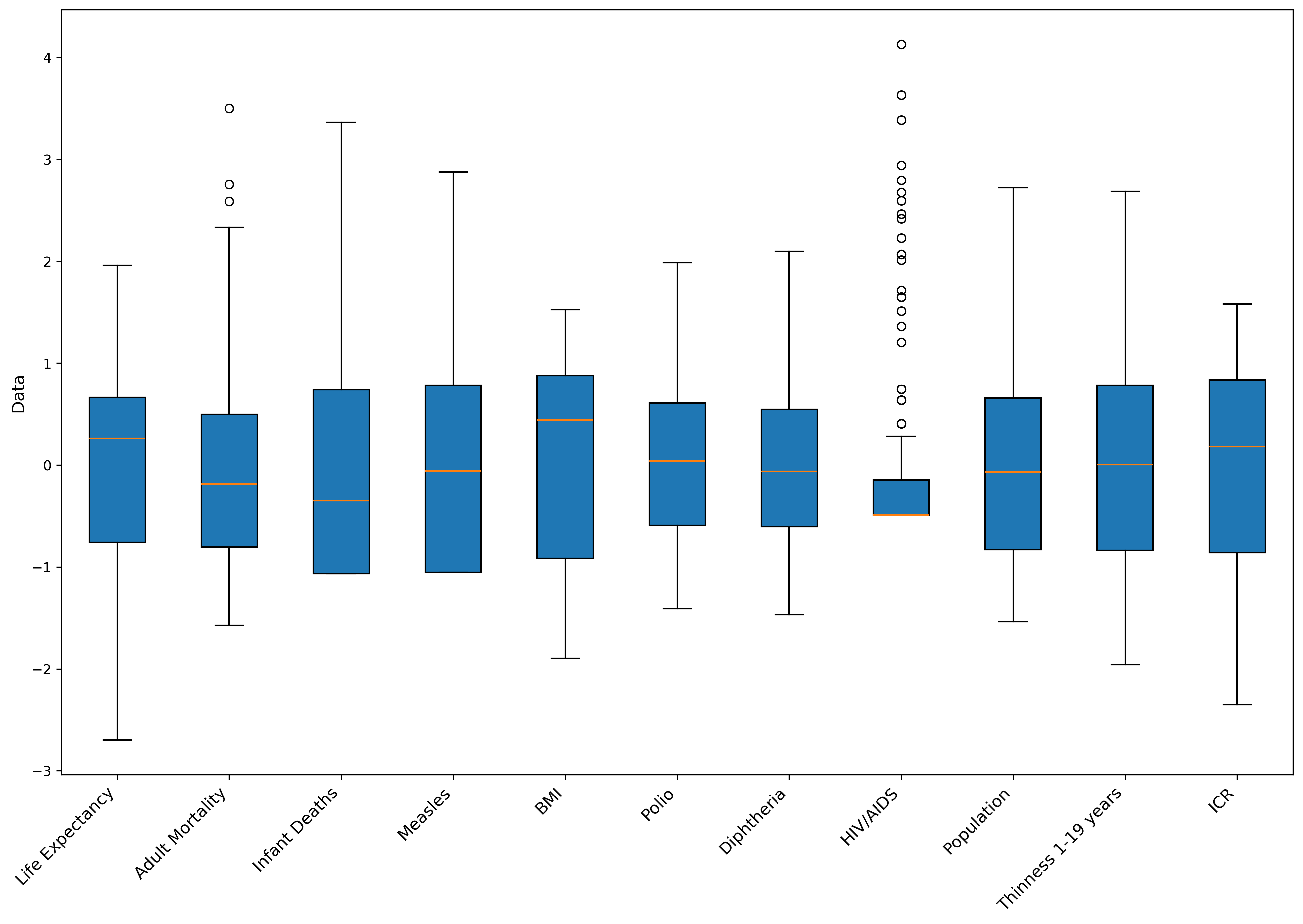}}
\caption{Box-plots of the standardized 2015-year dataset with sample size $n=141$.}
\label{fig:outlier_trans_stand_data2}
\end{figure}

Figures \ref{fig:outlier_trans_stand_data1} and \ref{fig:outlier_trans_stand_data2} show the Box-plots for the 2000-year and 2015-year standardized datasets. The median confirms the absence of skewness in the data where it is almost centered for most of the variables suggesting the symmetry. Even though some outliers appear in Adult Mortality and HIV/AIDS all observations are retained as they are real observations.      

\begin{table}[hbt!]
\scriptsize
\setlength{\tabcolsep}{1.2pt}
\renewcommand{\arraystretch}{0.7}
\begin{threeparttable}
\caption{Results of the trained ANN model against the other methods for the 2000-year and 2015-year datasets}
\label{tab:ex_results}
\begin{tabular}{l*{12}{l}}
\toprule
 & \multicolumn{6}{c}{\textbf{2000}} & \multicolumn{6}{c}{\textbf{2015}} \\
\textbf{Variable} 
& \textbf{ANN} & \textbf{Lasso} & \textbf{Forward} & \textbf{Backward} & \textbf{AIC} &\textbf{BIC}
& \textbf{ANN} & \textbf{Lasso} & \textbf{Forward} & \textbf{Backward} & \textbf{AIC} &\textbf{BIC} \\
\midrule
Adult Mortality 
& No & Yes & Yes & Yes & Yes & Yes
& Yes & Yes & Yes & Yes & Yes & Yes \\
Infant Deaths 
& No & No & No & No & No & No 
& No & No & No & No & No & No \\
Measles
& No & Yes & Yes & Yes & Yes & No 
& No & No & No & No & No & No \\
BMI 
& No & Yes & No & No & No & No 
& No & No & No & No & Yes & No \\
Polio 
& No & Yes & No & No & No & No
& No & Yes & Yes & Yes & Yes & Yes \\
Diphtheria 
& No & No & No & No & No & No 
& No & Yes & No & No & No & No \\
HIV/AIDS 
& Yes & Yes & Yes & Yes & Yes & Yes 
& Yes & Yes & Yes & Yes & Yes & Yes \\
Population (millions)
& No & Yes & No & No & No & No 
& No & Yes & No & No & Yes & No \\
Thinness 1-19 years
& No & Yes & Yes & Yes & Yes & No 
& Yes & Yes & Yes & Yes & Yes & Yes \\
ICR 
& Yes & Yes & Yes & Yes & No & Yes
& Yes & Yes & Yes & Yes & Yes & Yes \\
\bottomrule
\end{tabular}
\end{threeparttable}
\end{table}

 Table \ref{tab:ex_results} shows the variables selected by each method for both the 2000 and 2015 datasets.  Note that for the 2000 dataset the ANN gives HIV/AIDS and ICR being the important variables.  Note that all other methods except AIC also selected these two variables as well.  However, LASSO selected all variables except Infant Deaths and Diptheria. Forward and Backward both agree on the variables Adult Mortality, Measles, HIV/AIDS, Thinness 1-19 years and ICR as being important. The AIC approach selected Adult Mortality, Measles, HIV/AIDS and Thinness 1-19 years.  In contrast, the BIC method selected Adult Deaths, HIV/AIDS and ICR as important.  In this case the ANN seems to be more conservative in that it chooses fewest variables.  For the 2015 dataset the ANN gives Adult Deaths, HIV/AIDS, Thinness 1-19 years and ICR being the important variables.  Note that all other methods also selected these two variables as well.  In addition, LASSO selected all variables except Infant Deaths, Measles and BMI. Forward and Backward both agree on the variables Adult Mortality, Polio, HIV/AIDS, Thinness 1-19 years and ICR as being important. The AIC approach selected all variables except Infant Deaths, Measles and Diphtheria.  In contrast, the BIC method selected Adult Mortality, Polio, HIV/AIDS, Thinness 1-19 years and ICR as important.  This dataset is interesting in that each selection method produces a different set of important variables on the two sets of data.  The ANN approach appears to select fewer variables than the other approaches and hence is the conservative option in this case.

\section{Conclusion}\label{sec.conclusion}

This work develops an AI framework for model selection in the linear model context.  By using a large number of synthetic datasets covering a wide range of regression coefficients and model error variances an ANN can be trained to determine which predictors are important in defining the response variable.  The method is competitive with existing variable selection methods such as LASSO, forward, backwards, stepwise, AIC and BIC.  Furthermore, the power to detect important variables based on the magnitude of the regression coefficient is competitive with the alternative methods.  However, the approach appears to be more conservative in that it selects fewer variables than the competing methods.  Furthermore, the scalability of the approach allows for a varying number of predictors to be handled by a single ANN.   One should also note that this approach is easily deployable as the ANN that can handle up to $p=100$ variables has already been trained and does not need to be trained again on each new installation.  The approach is novel in that it uses Artificial Neural networks and synthetic datasets to make a pretrained AI that can simultaneously assess all variables importance.  Ideas of attention and multihead attention may be useful for situations where $n<p$ and in other large data cases to better assess across thousands of variables simultaneously.

The approach has been demonstrated on a real-world dataset which highlights not only the performance of the ANN but also several alternative approaches.  From this application, it is interesting that each method selected different predictors when using the same dataset. Again, the ANN approach seemed to be more conservative in this setting as well by selecting a fewer number of variables.  This may be related to the approach used to generate the training data with a small gap between -0.01 to 0.01 to avoid having a zero coefficient on a variable that is designed to be important.  Studies should be done on ways to improve performance on small regression coefficients to lower the false negative rate. 

As this is the first work of it's kind the potential for future work is great.  Simple extensions of the approach to handle missing data, categorical variables, multicollinearity, nonlinear regression, and multivariate regressions seem to be direct.  This approach could also be extended to Dynamic Models, time series models and combined with other feature selection methods.

\bibliographystyle{imsart-nameyear} 
\bibliography{AI.bib} 

\end{document}